\documentclass[10pt,twoside,twocolumn]{article}
\usepackage{graphicx} 

\begin{document}




\twocolumn[
  \begin{@twocolumnfalse}
\noindent\LARGE{\textbf{Human olfactory receptor 17-40 as active part of a nanobiosensor: A microscopic investigation of its electrical properties$^\dag$}}
\vspace{0.6cm}

\noindent\large{\textbf{Eleonora Alfinito,$^{\ast}$\textit{$^{a}$} Jean-Francois Millithaler,\textit{$^{a}$} Lino Reggiani,\textit{$^{a}$} Nadia Zine,\textit{$^{b}$} and
Nicole Jaffrezic-Renault\textit{$^{b}$}}}\vspace{0.5cm}


\vspace{0.6cm}

\noindent \normalsize{Increasing attention has been recently devoted to protein-based nanobiosensors. 
The main reason is the huge number of possible technological applications,
going from drug detection to cancer early diagnosis.
Their operating model is based on the protein activation and the corresponding conformational change, 
due to the capture of an external molecule, the so-called ligand.
Recent measurements, performed with different techniques on human 17-40 olfactory receptor, evidenced a 
very narrow window of response in respect of the odour concentration. 
This is a crucial point for understanding whether the use of this olfactory receptor as sensitive part of a nanobiosensor is a good choice.
In this paper we  investigate the topological and electrical properties of the human olfactory receptor 17-40 with the objective
of providing a microscopic interpretation of available experiments. 
To this purpose, we model the protein by means of a graph able to capture the mean features of the 3D backbone 
structure. The graph is then associated with
an equivalent impedance network,  able to evaluate  the impedance spectra of the olfactory receptor, in its native and activated state. We assume a topological origin of the different protein electrical responses to different ligand concentrations: In this perspective all the experimental data are collected and interpreted satisfactorily within a unified scheme, also useful for application to other proteins.}
\vspace{0.5cm}
 \end{@twocolumnfalse}
  ]



\footnotetext{\textit{$^{a}$~Dipartimento di Ingegneria dell'Innovazione, Universit\`{a} del Salento, Lecce, Italy, EU, 
\& CNISM - Consorzio Nazionale Interuniversitario per le Scienze Fisiche della Materia, Roma, Italy, EU. \\E-mail:eleonora.alfinito@unisalento.it}}
\footnotetext{\textit{$^{b}$~University of Lyon, Laboratory of Analytical Chemistry, Claude Bernard University,
43 Boulevard 11 Novembre 1918, 69622 Villeurbanne Cedex, France, EU}}




\section{Introduction}
Olfaction is one of the most refined senses in mammals: 
It allows to recognize many thousands of different odours.
In general, each odour is a mixture of different molecules, each of them able to activate a specific set of olfactory receptors (ORs), present in the nasal epithelium \cite{Gopel,Liu,Duchamp,Pardo}. 
By contrast, each OR selectively recognizes few specific molecules (ligands) \cite{Jacquier06}.
In such a way, the sequence of activated ORs identifies the odour univocally \cite{Jones,Triller08}. 
This mechanism indicates the ORs as the natural candidates for the production of a new generation of hybrid nanodevices for the detection of specific ligands
 \cite{Bond,Sab,physreve,Epl,Vestergaard,Gardner}. 
To this purpose, these nanobiosensors should integrate animal ORs  with an electrical/electronic transducer \cite{Bond}.

Conventional electronic noses are biomimetic sensors able to detect specific odours. The detection is usually performed by means of solid-state gas sensors:
They translate the captured information into a digital signal which is elaborated by appropriate algorithms.  To date,  the sensitivity and selectivity of conventional devices are of good level but still very far from those of a mammalian nose, the gold standard of an olfactory biosensor \cite{Pardo,Sab}.  The use of ORs as active part of a sensor aims to improve its selectivity, to reduce its size for achieving an on-field detection, to cut the costs and, as a future challenge, to obtain for each odour of interest a specific device \cite{Gopel,Liu,Bond,Pardo,Sab,Vestergaard,Gardner}.

These preliminary remarks are mainly supported by the observation of a change of the electrical response in an OR-based devices \cite{Benilova,Hou,Marra}. This change coincides with the protein activation due to the presence of the specific ligand.

\par
All the ORs are G-protein coupled receptors (GPCRs): This family of proteins
includes receptors of many different molecules and also the light receptors
(opsins). GPCRs have structure and function quite similar.
Previous  studies on opsins \cite{Sab} and rat OR I7 \cite{Sab,Jap} validated a theoretical model able to describe the electrical properties of these proteins through an impedance network protein analogue (INPA).   
Recent investigations on human OR 17-40 \cite{Benilova,Vidic,Levasseur},
valuable for a practical realization of a nanobiosensor,  led to interesting results  not yet fully understood on a microscopic ground. 
In particular, measurements were performed with electrochemical impedance spectroscopy (EIS) on a two terminal device with the active part controlled by this OR: It was observed that the polarization resistance showed a peculiar bell-shaped behaviour, at increasing values of the specific odorant concentration \cite{Benilova}. The peak was  centered at low concentrations of the specific ligand (around $10^{-10}$~M). 
Furthermore, by using complementary techniques like the differential surface
plasmon resonance (SPR) and the differential bioluminescence response, in addition to the first peak,  a second maximum response was observed at a higher concentration of the odorant (around $10^{-5}$  M)\cite{Vidic}.
These results were also confirmed by differential conductance measurements \cite{Marra}.

\par
The aim of this paper is to group recent investigations performed on human OR 17-40, in particular those obtained from EIS measurements, into a unified microscopic model.
By using the INPA approach we correlate the electrical response of the given protein to its topological properties. 
Accordingly, it is possible to predict  how the modifications of the latter, as induced by the capture of ligands, influence the former. 
The understanding of this correlation is promising for a future use of human OR 17-40 in nanobiosensors.  
Finally, we conjecture the existence of a correspondence between the \textit{collective} response of an ensemble of proteins to a specific odour concentration, and the variation of the calculated \textit{single} protein impedance at different values of the network free parameter $R_c$. This parameter defines the maximal distance between two interacting amino acids. 
In this  way, the single protein model is able to  provide  estimations on the electrical response of a protein based macroscopic device.
\par
The paper is organized as follows. 
Section \ref{sec:theory} briefly summarizes the theoretical model.
Sections \ref{sec:methods} and \ref{sec:results} detail the main results of the paper and  illustrate a conjecture for describing the protein modifications due to different ligand concentrations. 
A conclusive section summarizes the major achievements and the perspectives of this research.  
\section{Theory}
\label{sec:theory}
The INPA approach, as detailed in previous papers \cite{Sab,Jap}, correlates
the  electrical and topological properties of a sensing protein. 
The main objective  of INPA is to predict the protein electrical modifications as induced by a change of its topology.
For the reader convenience, its main features are here briefly recalled. 
The approach is structured in two steps.

The first step produces the protein analogue graph.
The input data of the model are the tertiary (3D) structures of the protein. 
These structures are  known from the protein data bank (PDB) available in literature \cite{PDB} or similarly (see, for example, Ref.[17]). Given the 3D structure, the protein is represented by means of a topological network (a graph). 
Each node of the graph corresponds to a single amino acid. 
If two amino acids are closer than an assigned interacting radius, $R_{c}$, then a link is drawn between the corresponding nodes.

The second step dresses the graph.
By associating an elemental impedance with each link, the graph becomes an impedance network. 
In other words, each link mimics a privileged channel for charge transfer and/or charge polarization \cite{Jap}.

The elementary impedance between nodes $i$ and $j$, $Z_{i,j}$, consists of an Ohmic resistance, $R$, parallely connected with a parallel plate capacitor, $C$, filled with a dielectric \cite{Sab,physreve,Jap}: 
\begin{equation}
Z_{i,j}={l_{i,j}\over {\mathcal{A}}_{i,j}}   
{1\over (\rho^{-1} + i \epsilon_{i,j}\, \epsilon_0\omega)}  
\label{eq:1}
\end{equation}                   
where ${\mathcal{A}}_{i,j}=\pi ({R_{c}}^2 -l_{i,j}^2/4)$, is the cross-sectional area between two spheres of radius ${R_{c}}$ centered on the $i$-th and $j$-th node, respectively;
$l_{i,j}$ is the distance between these centers, $\rho$ is the resistivity, taken to be the same for every amino acid, with the indicative value of $\rho = 10^{10}$~$\Omega$m; $i=\sqrt{-1}$ is the imaginary unit, $\epsilon_0$ is the vacuum permittivity, and  $\omega$ is the circular frequency of the applied voltage. 
The relative dielectric constant of the couple of $i,j$ amino acids, $\epsilon_{i,j}$, is expressed in 
terms of the intrinsic  polarizability of the $i,j$ amino acids.
\par
By positioning the input and output electrical contacts, respectively on the first and last node (corresponding to the first and last amino acid of the protein sequential structure), the network is solved within a linear Kirchhoff scheme \cite{Sab,physreve,Jap}.
Within this framework, the protein is mapped into an electrical network whose topology reflects the protein conformation.
When the conformation changes, the network topology and, in turn, the electrical response, follows this change.
\section{Material and Method}
\label{sec:methods}
To date, no crystallographic data are available for the 3D structures of the human OR 17-40.
Accordingly, the 3D structures used in this paper are obtained from an homology modeling  \cite{Jap}. 
Those are calculated by using the light receptor bovine rhodopsin in the native state (in dark), as template for the henceforth mentioned native human OR 17-40, and the bovine rhodopsin in the active state, metarhodopsin II (in light), as template for the henceforth mentioned active human OR 17-40.
\par
The INPA approach is very flexible, and allows one to carry out  different kinds of investigations. 
In particular, here we point: 
\par
i)  To compare the contact maps of the protein, in the native and active state, as a function of the interacting radius, $R_c$.
These 2D maps visualize the  connected parts of the network, and are drawn by assigning to each couple of linked amino acids, say $i-j,$ a point of coordinates $i,j$. Two amino acids are linked if their distance is less than $R_c$, otherwise they remain separated.  
Therefore,  by comparing the contact maps of the protein in the native and active state we  gain access to a direct image of the main changes in the protein topology. 
\par
ii) To calculate the relative variation of resistance between the native and active state of the OR. 
Since  the elementary impedance depends on the distance between  amino acids, when the protein conformation changes also the global impedance, and thus the resistance, changes.
\par
iii) To calculate, within the standard frequency range 1~mHz~$\div$~100~kHz, the protein impedance spectra for the native and active states. In this way it is possible to draw the theoretical Nyquist plot that should be compared with the experimental one. 
\par
Finally, the different electrical responses, as \textit{calculated} from the networks pertaining to the native/active state, are here taken as an estimate of the \textit{measured} electrical responses implied by the protein conformational change.  
\section{Results}
\label{sec:results}
The main results concerning sections \ref{sec:methods} and \ref{sec:theory} are detailed in the following subsections.
\subsection{Protein topology}
\begin{figure}[h]
\centering
  \includegraphics[height=6cm]{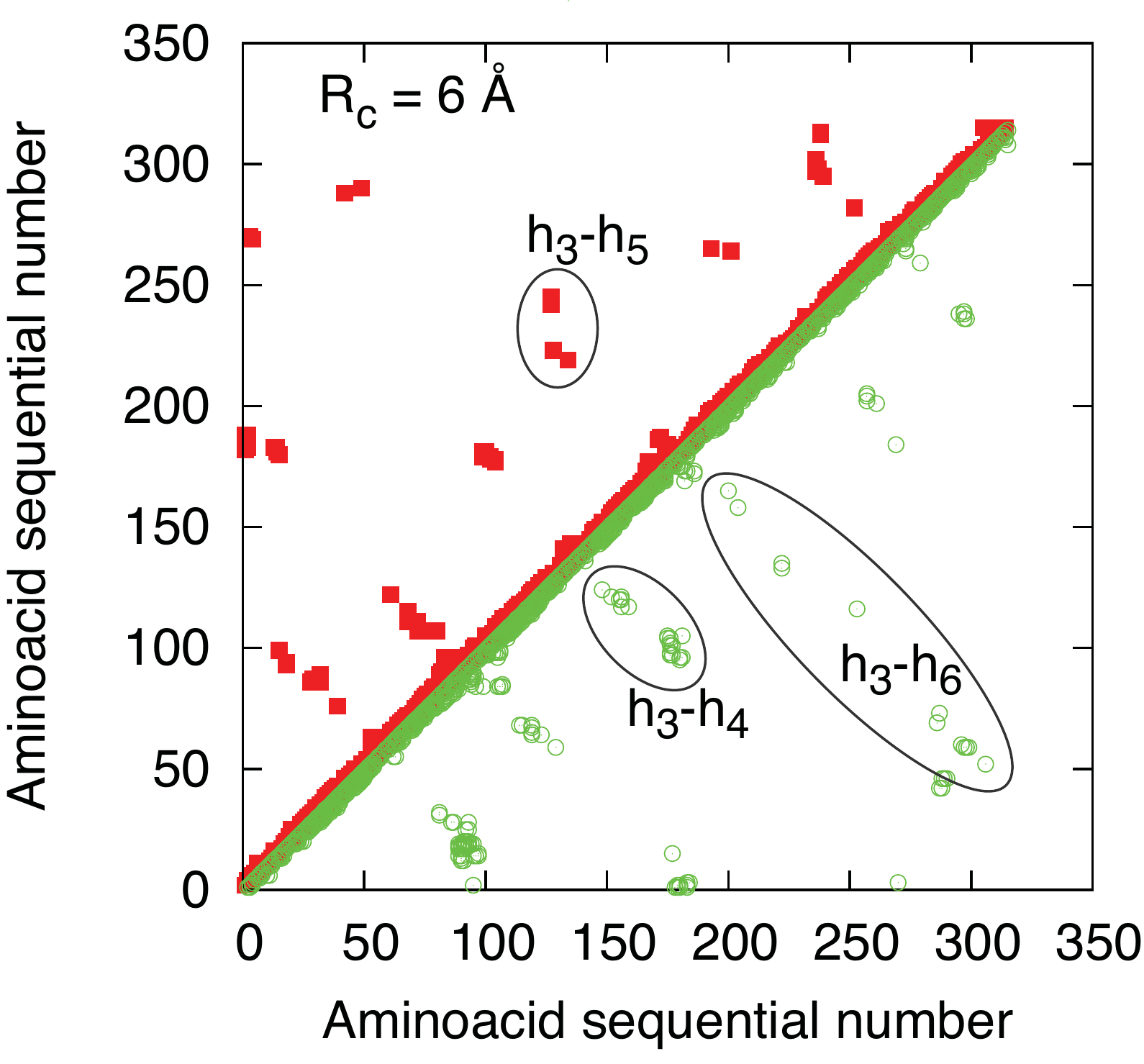}
  \caption{(Color on line) Contact maps of human OR 17-40  for $R_{c}$= 6 \AA. The data of the native state are reported on the left side of the diagonal (black full squares), the data of the active  state are reported on the right side of the diagonal (green open squares). The ellipses signal the main differences between the native and active state which are attributed to the transmembrane helices h3-h4-h6.}
  \label{fig:1}
\end{figure}

\begin{figure}[h]
\centering
  \includegraphics[height=6cm]{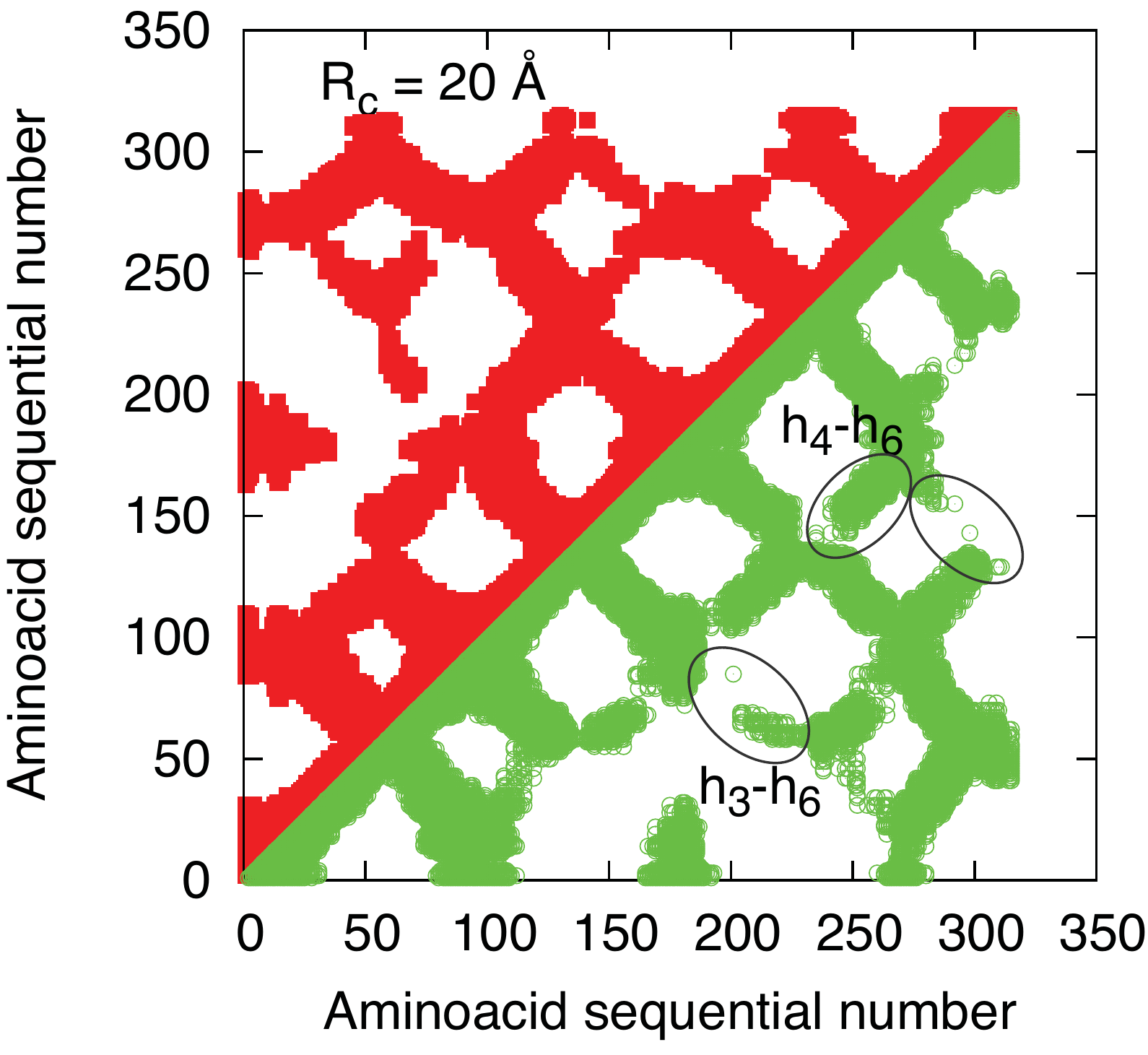}
  \caption{(Color on line) Contact maps of OR 17-40 for $R_{c}$= 20 \AA. The data of the native state are reported on the left side of the diagonal (black full squares), the data of the active state are reported on the right side of the diagonal (green open squares). The ellipses signal the main differences between the native and active state which are attributed to the transmembrane helices h3-h4-h6.}
  \label{fig:2}
\end{figure}

The global insight on the protein conformational change, as induced by the ligand capture, is given by the contact maps reported in Fig. 
\ref{fig:1} and Fig. \ref{fig:2}.
In these figures the diagonal is a symmetry axis that allows one to
compare native and active states in the same figure, the native state is represented by points satisfying $i<j$ while the active state by points $i>j$. 
Furthermore, the contact maps are calculated for two different interaction radii, specifically 6~\AA\, in Fig. \ref{fig:1} and 20~\AA\, in Fig. \ref{fig:2}. 
From these figures we can observe that, by increasing the value of $R_c$, the density of points and thus the connectivity increase substantially, as expected.
In particular, the main differences between the native and active state are in the increased connectivities among the helices h3-h4-h6, where the principal binding pocket is presumably located \cite{Hall}. 
Notably, significant differences between native and active states survive at large $R_{c}$, say up to $R_c$ values of about 50 \AA.

%
\subsection{Protein resistance}
%
%
\begin{figure}[h]
\centering
     \includegraphics[width=7.cm]{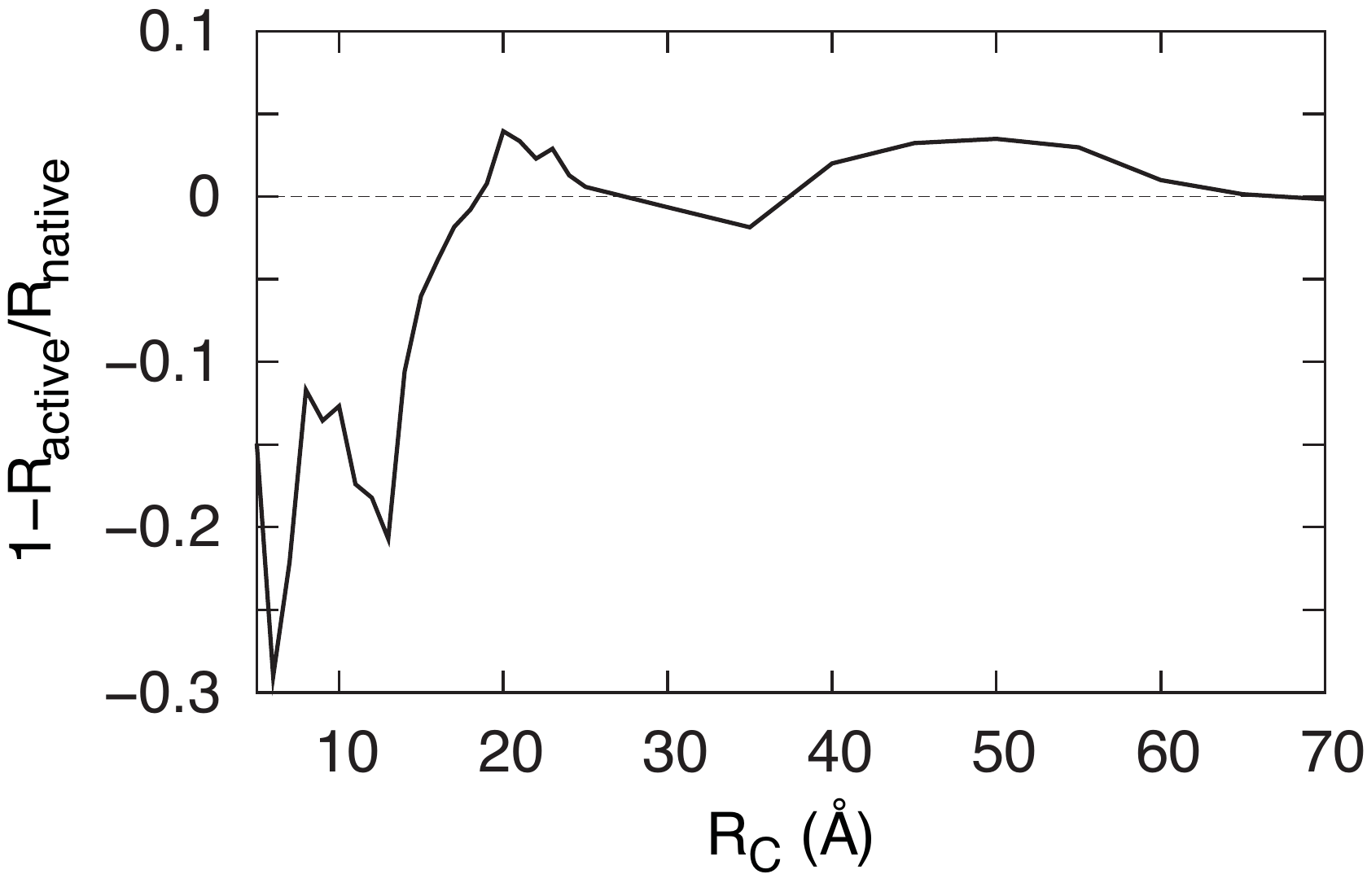}
     \caption{Relative resistance variation of human OR 17-40 as function of the interacting radius $R_{c}$ .}
     \label{fig:3}
\end{figure}
The value of the protein resistance in its native state is in general different from that in its active state, and the magnitude of this difference depends on $R_{c}$ as reported in Fig. 3. 
Here the main result is the non-monotonic behaviour of the resistance change at increasing values of $R_c$. 
The region of maximum sensitivity to the conformational change is found for  $R_{c}$ in the range 6 $\div$ 14 \AA.
Furthermore, in the ranges of $R_c$ values 18 $\div$ 26 \AA\, and  38 $\div$ 65 \AA,  calculations evidence an inversion of the resistance variation, with the active state becoming less resistive than the native state.
In terms of the electrical network, such an inversion is interpreted  as a stronger increase of parallel with respect to series connections of the active link resistances. 
\par
Recent advances on the activation mechanism of these proteins \cite{Kolbika} depict their dynamics in terms of a complex series of intermediate conformational transitions (that involve the disruption of non covalent intramolecular interactions) rather than in terms of a simple on/off switch. 
These interactions stabilize the protein state in a specific equilibrium condition, which depends on the  ligand concentration.  
An increasing concentration of ligands induces the disruption of stabilizing interactions, enabling the receptor to evolve toward a more active state.
%
%
\begin{figure}[h]
\centering
     \includegraphics[width=7.cm]{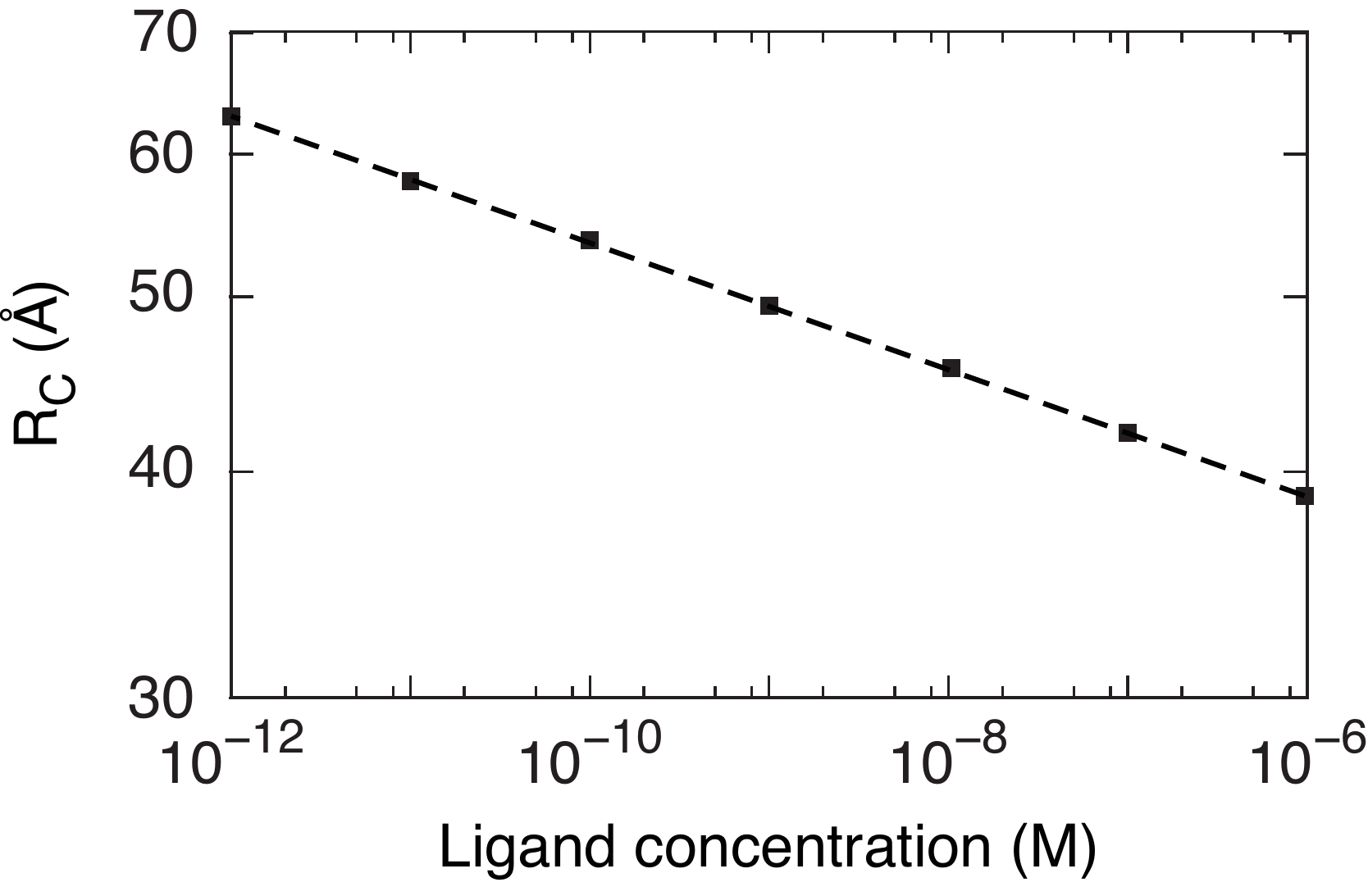}
     \caption{Interacting radius ($R_{c}$) as function of the specific ligand helional concentration, $\Lambda$.
Dashed line is a guide to the eye.}
     \label{fig:4}
\end{figure}
\par
In our model, we conjecture that the disruption of stabilizing interactions is related to a reduction of the interaction radius $R_c$, which corresponds to a reduction of connections between amino acids. 
Accordingly, we assume the existence of a correlation between the interacting radius, $R_c$, and the concentration of a specific odorant, $\Lambda$. 
In this way, the change of the single protein resistance  vs $R_c$ (see Fig. 3) is correlated with the analogue quantity measured for different ligand concentrations \cite{Benilova}.
In establishing this correspondence we assume that the peak of the electrical response of human OR 17-40 (occurring at $\Lambda= 10^{-10}$ M) corresponds to $R_{c}$= 52 \AA, as resulting by Fig.~3. 
Furthermore, in this perspective, Fig.~3 gives the evolution of the relative resistance variation at different ligand concentrations.
Starting from very large $R_c$, where
there is no difference between state responses (small $\Lambda$ values in our conjecture), at smaller $R_c$ the resolution of the electrical response 
evolves toward larger values  (large $\Lambda$ values). 
In doing so, the resolution exhibits at least two bumps. 

The functional dependence of $R_{c}$ on $\Lambda$ is found  to  follow a power law, as evidenced by the dashed line in Fig. \ref{fig:3}.
To enforce the above conjecture, Fig. \ref{fig:4}  reports $R_{c}$ as a function of a specific odorant concentration.
Here, the data are obtained by comparing  the experimental  response for different helional concentrations \cite{Benilova} with the theoretical relative resistance variation of the single protein at different $R_{c}$ values, as reported in Fig. \ref{fig:3}.

Figure \ref{fig:5} reports the polarization resistance variation as a function of the odorant concentration, as measured  for the specific cases of heptanal and helional \cite{Benilova}. 
Theoretical results (dashed curve)  are found to reproduce qualitatively  well the experiments (symbols), which evidence  a typical bell-shaped behaviour with a maximum sensitivity for the case of helional at a molar concentration of $10^{-10}$ M.
By construction,  the quantitative comparison between theory and experiments cannot discriminate the different sensitivity  to different odorants of the OR.
In any case, at this stage of investigation we believe that the overall agreement between theory and experiments is satisfactory enough to further  support  our conjecture.
\par
Finally, we notice that the dose-response depicted in Fig. 5 departs from the saturation behaviour observed for other receptors (like rat OR I7) where an increasing concentration of the specific ligand produces a systematic increasing of the protein response \cite{Hou}. 
According to our conjecture,  the origin of this difference should be found in the peculiar modification of the topology  undergone by  human OR17-40.
\subsection{Protein impedance spectrum} 
%
%
\begin{figure}[h]
\centering
     \includegraphics[width=6cm]{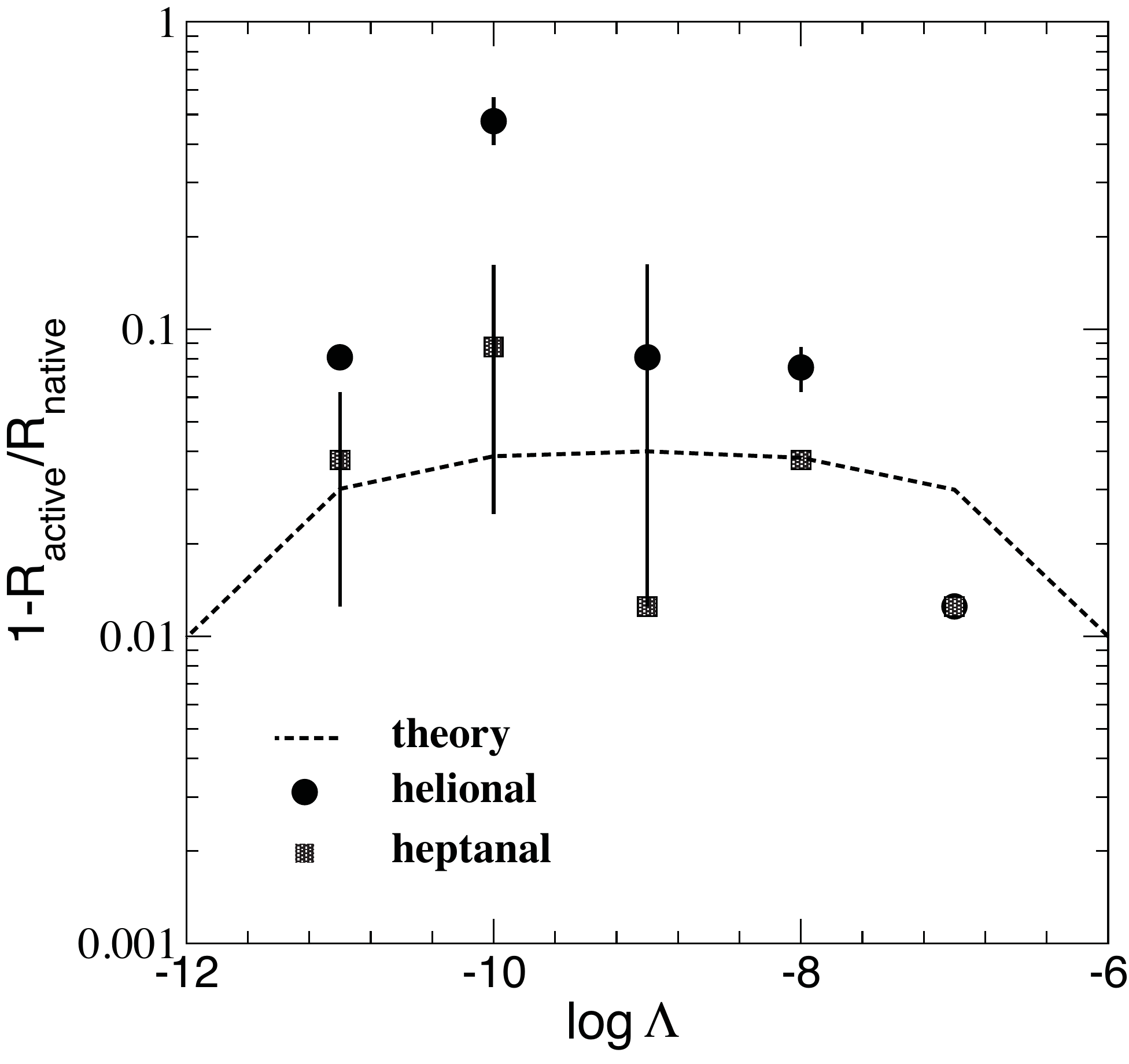} 
     \caption{Per cent change of the polarization resistance at increasing concentration of specific odorants for  human OR 17-40 at room temperature.
Symbols refer to experiments, with bars giving the estimated uncertainty~\cite{Hou,Benilova}, dashed line refer to theoretical expectations.}
     \label{fig:5}
\end{figure}
%
\begin{figure}
\centering
      \includegraphics[width=.45\textwidth]{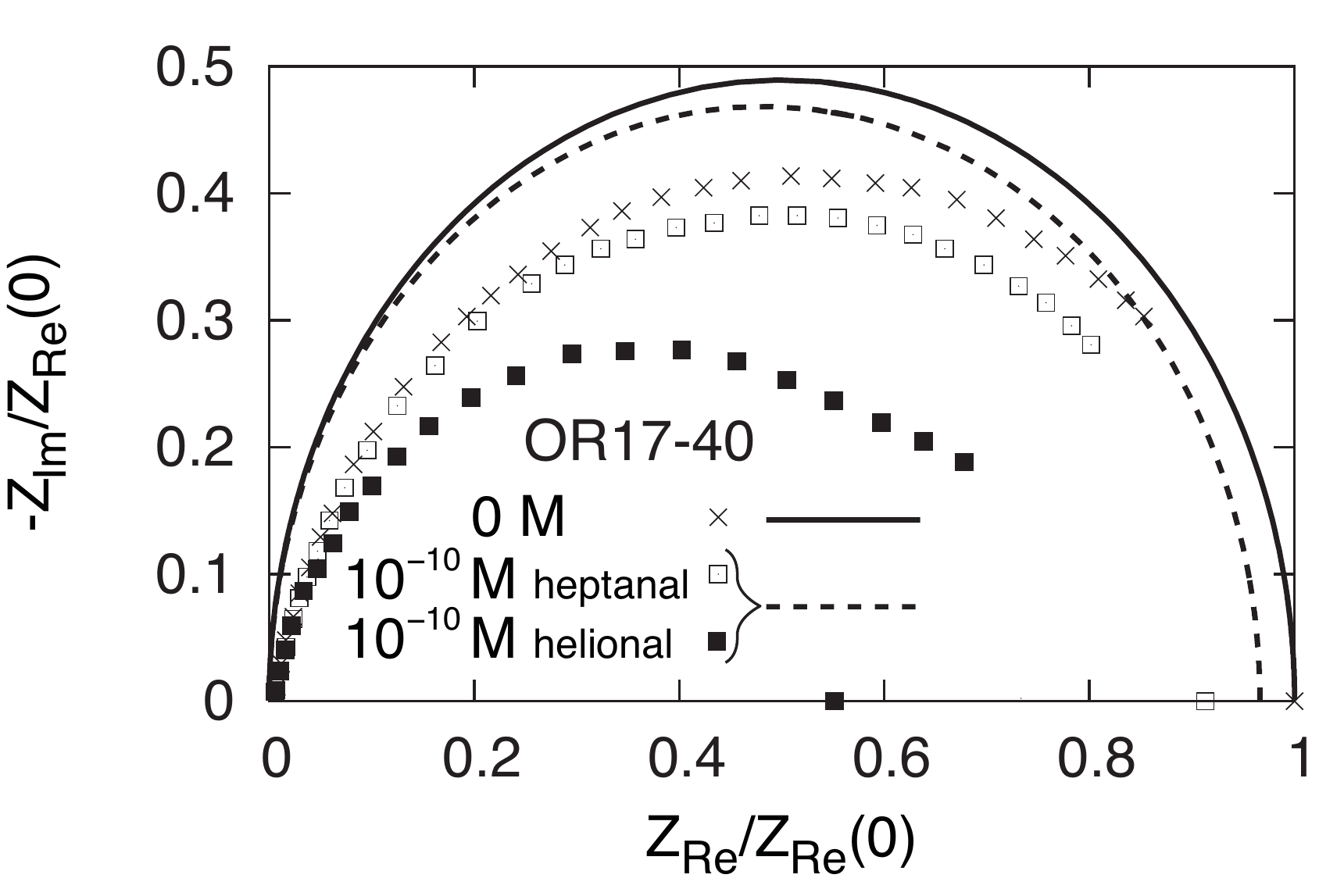}
\caption{Nyquist plot of human OR 17-40 in the absence and in the presence of a specific ligand (heptanal, helional). The impedances are normalized to the static value of the native state, $Z_{Re}^{nat}(0) = 33 \rm{\ K \Omega \ cm^2}$. Symbols pertain to experiments with:  Crosses referring to no odorant,  empty (full) squares to  an heptanal (helional) concentration of $10^{-10}$  M  at room temperature ~\cite{Hou,Benilova}. Curves pertain to theoretical results with: Continuous curve referring to the native state configuration as input data with $R_{c}= 70$\AA,  dashed line referring to the activate state configuration with $R_{c} = 46$ \AA. }
\label{fig:6}
\end{figure}
The impedance spectrum of the single protein is  explored over a wide range of frequencies  and the results are given by means of the associated Nyquist plot. 
This plot is obtained by drawing the negative imaginary part versus the real part of the global impedance, within a given frequency range (typically from 1 mHz to 100 kHz as in experiments  \cite{Benilova}). 
Figure ~\ref{fig:6} reports the Nyquist plots of human  OR 17-40 with the impedance normalized to the static value of the native state. 
Symbols pertain to experiments, with crosses referring to the absence of a specific odorant, full (empty) squares  to the presence of the specific odorant helional (heptanal) at the concentration  reported in the figure.  
Curves pertain to theoretical results where the single protein is taken to be representative of the entire sample, and with continuous (dashed) lines referring to the native (active) state.
Within our model, the Nyquist plots are obtained by using as input data the networks corresponding to the native and active states at the cut-off radius which, according to Fig.~\ref{fig:5}, gives the maximum resolution. 
The agreement between theory and experiments is found to be satisfactory from a qualitative point of view and, apart for a significant underestimation of the maximum experimental value,  acceptable from a quantitative point of view.
We remark that the near ideal semicircle shape of the experimental  Nyquist plot is well reproduced, thus confirming that the network impedance model behaves closely to a single RC circuit as expected by the presence of a rather uniform distribution of time constants associated with the different values of the resistance and capacitance of the links \cite{Jap}.
The scarcity of experimental data and the lack of a well certified knowledge for the 3D structures of the considered proteins lead us to consider these results as a first but significant step towards a microscopic modeling of the electrical properties  of this olfactory receptor.    
\section{Conclusions and perspectives}
Recent EIS experiments on human OR 17-40  showed that 
the mechanism of odorant capture can be monitored by means of the modification of the impedance spectra \cite{Benilova}.
These results are generally confirmed by surface plasmon resonance and bioluminescence response of appropriately prepared OR 17-40 samples \cite{Benilova,Vidic}.
The EIS results are here microscopically interpreted on the basis of the conformational change of the protein tertiary structure, induced by the sensing action. 
Accordingly, the conformational change induces a variation  of the protein electrical response that is detectable with the EIS technique and can be reproduced by means of the INPA modeling.
In particular, a possible correlation between the interaction radius, $R_c$, at the basis of charge transfer between amino acids, and the odorant concentration, $\Lambda$, is inferred by comparing theory with experiments.

From one side, the present INPA approach is rather essential and the qualitative agreement with available experiments is used as a proof of concept that is promising for the realization of an electronic nose based on an array of nanobiosensors. 
Furthermore, the narrow window of sensitivity of human OR 17-40 to the odorant concentration poses potential  limitations in the advantages of receptor based biosensors in respect of traditional sensors. These limitations should be overcome by investigating similar ORs which are, like rat OR I7 \cite{Jap}, potentially  free from these drawbacks.  

From another side, relevant information on the protein tertiary structure and the mechanisms of charge transfer can be extracted from  the microscopic interpretation of experiments.
The next step is the production of new identified ORs in efficient heterologous systems (mammalians or yeast cells), to evaluate their functional response. 
In  parallel, improvements of the theoretical  model should  take into account the role of temperature, hydration, and other details.

\vspace{0.5cm}
This research is carried out within the bioelectronic olfactory neuron device (BOND) project sponsored by the CE within the 7th Program, grant agreement: 228685-2.
Dr V. Akimov and the \textit{INRA UR1077 MIG, Jouy-en-Josas, France} are gratefully acknowledged for providing the 3D structures of the OR 17-40. 
Dr. E. Pajot (INRA UR1197 NOeMI, Jouy-en-Josas, France) is thanked for useful discussions on the subject.



\end{document}